\documentclass[aps,prb,reprint]{revtex4-2}

\usepackage{graphicx}

\usepackage{bm}

\begin{document}

\title{Suppression of orbital ordering and emergence of a glassy magnetic state in a high-entropy spinel vanadate}

\author{Shun Ito}
\author{Sota Nakakuki}
\author{Satoshi Demura}
\author{Tadataka Watanabe}
\email{watanabe.tadataka@nihon-u.ac.jp}
\affiliation{Department of Physics, College of Science and Technology, Nihon University, Chiyoda, Tokyo 101-8308, Japan}
\date{\today}

\begin{abstract}
High-entropy oxides provide a unique platform for exploring the interplay between configurational disorder and correlated electronic states. We report on structural, thermodynamic, and magnetic properties of the high-entropy spinel vanadate (Li$_{0.2}$Mg$_{0.2}$Mn$_{0.2}$Co$_{0.2}$Zn$_{0.2}$)V$_2$O$_4$, in which orbital-active V ions occupy the pyrochlore sublattice. X-ray diffraction measurements reveal that the cubic spinel structure is preserved at temperatures down to 5 K without any detectable symmetry lowering. Unlike conventional spinel vanadates exhibiting a symmetry-lowering structural transition driven by orbital ordering, the compound that we study exhibits a suppression of long-range orbital ordering in the high-entropy state. Magnetic susceptibility measurements show glassy magnetic freezing at temperatures below $\sim$20 K, while specific-heat measurements reveal no anomaly associated with long-range ordering down to 3 K. Frequency-dependent ac susceptibility reveals pronounced glassy dynamics. Analyses based on the Mydosh parameter, dynamic scaling law, and Vogel--Fulcher law suggest an intermediate dynamical regime between canonical spin-glass and cluster-glass behavior. Furthermore, Cole--Cole analyses reveal systematic deviations from a single Debye relaxation process, indicating the presence of heterogeneous magnetic relaxation dynamics near the freezing regime. Our results demonstrate that, in a frustrated spinel vanadate, high configurational entropy suppresses long-range orbital ordering and stabilizes a heterogeneous glassy magnetic state, highlighting the combined roles of geometrical frustration, orbital degrees of freedom, and configurational disorder in high-entropy magnets.
\end{abstract}

\maketitle

\section{Introduction}

Geometrically frustrated magnets have attracted considerable attention because competing interactions in these systems can suppress conventional long-range ordering, giving rise to a variety of unconventional ground states \cite{Lacroix}. Spinel oxides $AB_2$O$_4$, in which magnetic ions on the $B$ site form a pyrochlore lattice consisting of corner-sharing tetrahedra, provide a prototypical platform for studying frustration. Representative examples include spinel chromites $A$Cr$_2$O$_4$, which exhibit spin-driven lattice distortions, and spinel vanadates $A$V$_2$O$_4$, for which spin, orbital, and lattice degrees of freedom are strongly coupled \cite{Lee}.

Spinel vanadates $A$V$_2$O$_4$ are particularly intriguing because the V$^{3+}$ ions possess both spin and orbital degrees of freedom. In these compounds, the two electrons occupying the $t_{2g}$ orbitals are subject to strong orbital degeneracy and geometrical frustration. $A$V$_2$O$_4$ with nonmagnetic $A$ = Zn, Mg, and Cd exhibits, upon cooling, a structural phase transition followed by an antiferromagnetic phase transition \cite{Ueda,Reehuis,Lee2,Zhang,Mamiya,Wheeler,Nishiguchi,Onoda,Giovannetti,Watanabe}. In contrast, for magnetic $A$ = Mn and Fe, additional $A^{2+}$-V$^{3+}$ exchange interactions give rise to richer structural and magnetic phase diagrams involving successive structural and ferrimagnetic transitions \cite{Adachi,Suzuki,Zhou,Chung,Garlea,Hardy,Nii1,Nii2,Gleason,Katsufuji,MacDougall,Zhang2,Kang,Kawaguchi}.

For $A$V$_2$O$_4$ ($A$ = Zn, Mg, Cd, Mn, and Fe), the structural phase transition is generally associated with long-range ordering of the V $t_{2g}$ orbitals, which partially lifts the magnetic frustration by lowering the crystal symmetry and modifying the exchange interactions. This orbital ordering has been discussed in terms of competing Jahn--Teller coupling, Kugel--Khomskii exchange interactions, and relativistic spin-orbit coupling \cite{Lee2,Garlea,Wheeler,Suzuki,Kugel,Tsunetsugu,Tchernyshyov,Matteo,Marita,Kaur,Sarkar,Pardo,Watanabe}. Consequently, for $A$V$_2$O$_4$, the interplay among spin, orbital, and lattice degrees of freedom plays a central role in determining low-temperature properties.

Among spinel vanadates, CoV$_2$O$_4$ exhibits unique behavior near the itinerant-electron limit \cite{Kismarahardja,Kiswandhi,Huang,Kaur2,Kismarahardja2,Ma,Koborinai,Reig-i-Plessis,Watanabe2}. Although a ferrimagnetic transition occurs at $T_C \sim$ 150 K, the nature of the orbital state remains controversial. Several studies have suggested the suppression of long-range orbital ordering and the emergence of orbital-glass-like behavior \cite{Koborinai,Reig-i-Plessis,Watanabe2}. CoV$_2$O$_4$ therefore provides an important example of a spinel vanadate in which long-range orbital ordering is destabilized without completely removing the orbital degrees of freedom. These observations indicate that long-range orbital ordering in spinel vanadates can be destabilized by competing interactions, itinerancy, and related perturbations.  The resulting orbital state is therefore expected to be sensitive to additional perturbations such as those caused by configurational disorder.

Recently, high-entropy oxides have emerged as a new class of materials in which multiple cations occupy an equivalent crystallographic site. Since the discovery of entropy-stabilized oxides in 2015, extensive effort has been devoted to understanding their structural, magnetic, electronic, and functional properties \cite{Rost,Oses}. Among the various high-entropy oxides, spinel oxides have attracted particular interest because both the tetrahedral $A$ site and octahedral $B$ site can accommodate a wide variety of cations, providing a highly flexible platform for tuning magnetic and electronic properties. Recent studies have demonstrated that the magnetic properties of high-entropy spinels can be substantially modified through cation selection \cite{Musico}. However, most studies have focused on magnetic and functional properties, while the influence of configurational disorder on orbital degrees of freedom remains largely unexplored.

Very recently, $A$-site-disordered high-entropy spinel chromites (h-$A$)Cr$_2$O$_4$ containing Ni and Cu were reported to undergo cubic-to-orthorhombic structural transitions and long-range antiferromagnetic ordering without the intermediate tetragonal Jahn--Teller phase observed in stoichiometric NiCr$_2$O$_4$ and CuCr$_2$O$_4$ \cite{Mandal}; here, (h-$A$) denotes a high-entropy spinel $A$ site. This result suggests that configurational disorder can strongly modify cooperative Jahn--Teller and magnetostructural transitions in spinel oxides. Nevertheless, the influence of configurational disorder on the orbital degrees of freedom of the frustrated V pyrochlore sublattice remains largely unexplored.

In a notable recent study of high-entropy perovskite vanadates (h-$R)$VO$_3$, the effect of rare-earth-site disorder on the long-range orbital ordering of V$^{3+}$ ions was investigated \cite{Yan}; here, (h-$R$) denotes a high-entropy rare-earth site. This study showed that long-range orbital ordering survives even in the presence of substantial configurational disorder, and that the orbital state is governed primarily by the average ionic radius and size variance of the rare-earth ions. These results suggest that orbital degrees of freedom can remain robust in high-entropy oxides, raising an important question as to how long-range orbital ordering evolves in more strongly frustrated systems.

In contrast to perovskite vanadates $R$VO$_3$, spinel vanadates $A$V$_2$O$_4$ combine orbital degeneracy, geometrical frustration, and strong exchange interactions on the pyrochlore lattice formed by the V ions. Furthermore, magnetic ions occupying the $A$ site introduce additional exchange pathways and disorder effects absent in perovskite systems. These characteristics make high-entropy spinel vanadates (h-$A)$V$_2$O$_4$ an ideal platform for investigating the interplay of orbital physics, geometrical frustration, and configurational disorder. Nevertheless, the fate of long-range orbital ordering in such high-entropy spinel vanadates remains largely unexplored.

In this work, we investigate the structural, thermodynamic, and magnetic properties of the high-entropy spinel vanadate (Li$_{0.2}$Mg$_{0.2}$Mn$_{0.2}$Co$_{0.2}$Zn$_{0.2}$)V$_2$O$_4$. Combining synchrotron x-ray diffraction, laboratory x-ray diffraction, and heat-capacity measurements, we show that the cubic spinel structure persists down to low temperatures without any detectable structural transition. Furthermore, dc and ac magnetic susceptibility measurements reveal that, in contrast to conventional spinel vanadates exhibiting long-range orbital ordering, the high-entropy spinel vanadate exhibits a glassy magnetic state accompanied by heterogeneous magnetic relaxation dynamics. Our results demonstrate that high configurational entropy suppresses long-range orbital ordering and stabilizes a glassy magnetic state with heterogeneous relaxation dynamics, providing a new avenue for realizing unconventional spin-orbital states in frustrated high-entropy oxides.

\section{Experimental}

Polycrystalline samples of the high-entropy spinel vanadate (Li$_{0.2}$Mg$_{0.2}$Mn$_{0.2}$Co$_{0.2}$Zn$_{0.2}$)V$_2$O$_4$ and, for comparison, the high-entropy spinel manganate (Li$_{0.2}$Mg$_{0.2}$Mn$_{0.2}$Co$_{0.2}$Zn$_{0.2}$)Mn$_2$O$_4$ were prepared using a conventional solid-state reaction method. Stoichiometric amounts of the corresponding starting materials (Li$_2$CO$_3$, MgO, MnO, CoO, ZnO, V$_2$O$_3$, and Mn$_2$O$_3$) were thoroughly mixed, ground, and pressed into pellets. The pellets of (Li$_{0.2}$Mg$_{0.2}$Mn$_{0.2}$Co$_{0.2}$Zn$_{0.2}$)V$_2$O$_4$ were sintered in evacuated quartz tubes at 1100 $^{\circ}$C for 24 hours followed by furnace-cooling, while those of (Li$_{0.2}$Mg$_{0.2}$Mn$_{0.2}$Co$_{0.2}$Zn$_{0.2}$)Mn$_2$O$_4$ were sintered in air at 1100 $^{\circ}$C for 32 hours and then quenched to room temperature. Powder x-ray diffraction (XRD) measurements were carried out using both laboratory and synchrotron radiation. Laboratory XRD measurements were carried out using a Rigaku Ultima IV diffractometer with Cu K$\alpha$ radiation at temperatures down to 5 K. Synchrotron XRD measurements were performed at the BL02B2 beamline of SPring-8 using x rays with a wavelength of 0.4960618 \AA{}  at temperatures down to 100 K. Rietveld refinements of the diffraction patterns were carried out using the Z-Rietveld program \cite{Oishi,Oishi2}. The microstructure and chemical homogeneity of the samples were examined using a scanning electron microscope (SEM, HITACHI S-2150) equipped with an energy-dispersive x-ray spectrometer (EDS, HORIBA EMAX-7000). The temperature dependence of the dc magnetization and ac susceptibility was measured using a superconducting quantum interference device (SQUID) magnetometer (Quantum Design MPMS) over the temperature range 3 K $\leq T \leq$ 300 K. The ac susceptibility was measured using an ac excitation field of 2 Oe under zero dc magnetic field. Specific heat measurements were carried out using a Quantum Design PPMS over the temperature range 3 K $\leq T \leq$ 200 K.

\begin{figure}[t]
\centering
\includegraphics[keepaspectratio,width=\linewidth]{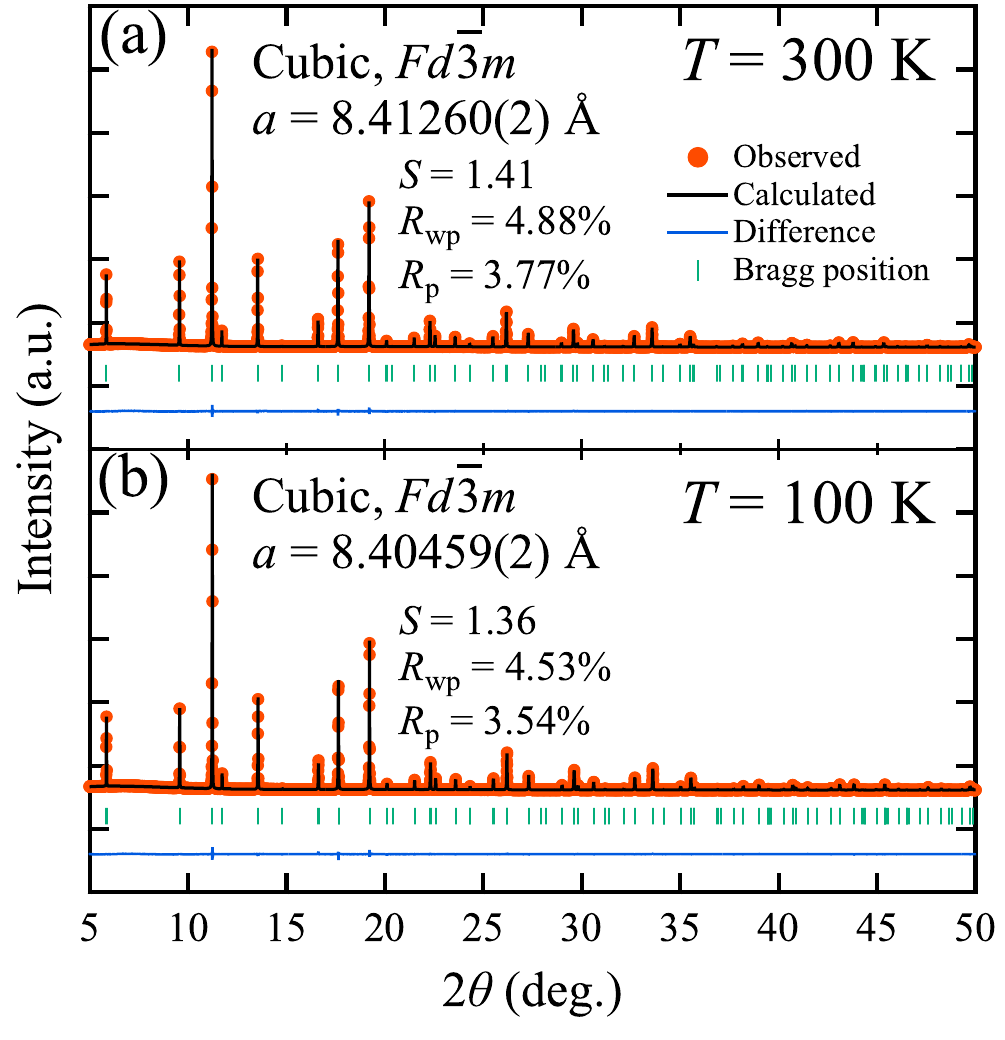}
\includegraphics[keepaspectratio,width=\linewidth,trim= 0 150 0 150]{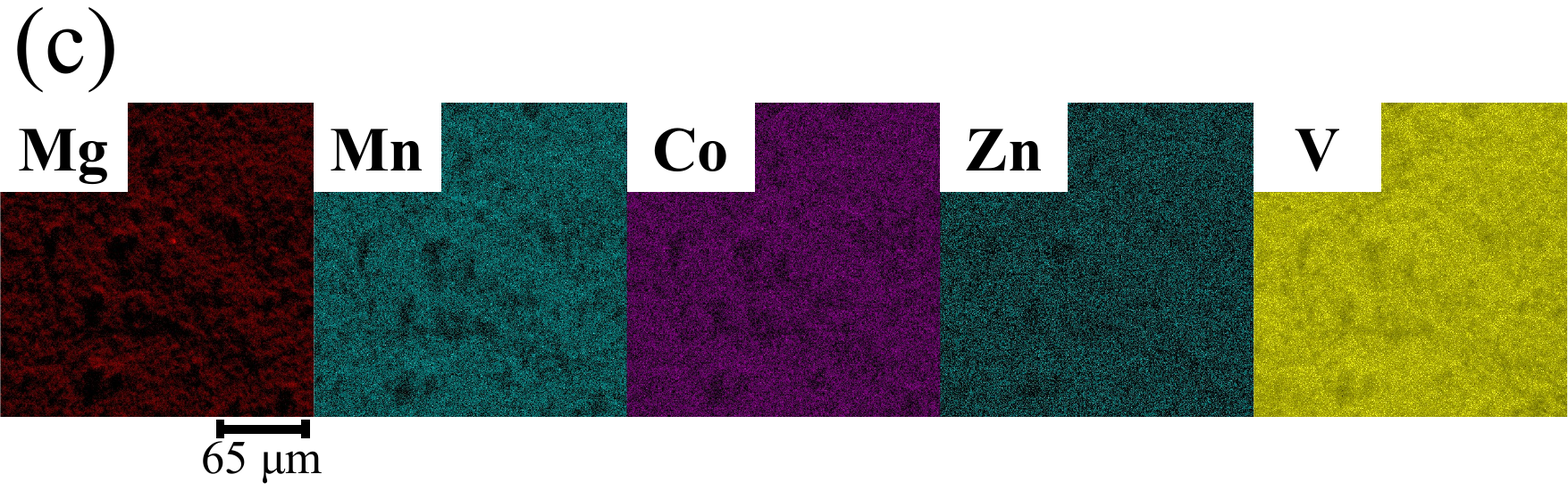}
\caption{Synchrotron XRD patterns of (Li$_{0.2}$Mg$_{0.2}$Mn$_{0.2}$Co$_{0.2}$Zn$_{0.2}$)V$_2$O$_4$ and the results of Rietveld analysis assuming the cubic space group $Fd\bar{3}m$: (a) 300 K and (b) 100 K. No peak splitting indicative of symmetry lowering from the cubic spinel structure is observed down to 100 K. (c) SEM image and EDS elemental mapping of (Li$_{0.2}$Mg$_{0.2}$Mn$_{0.2}$Co$_{0.2}$Zn$_{0.2}$)V$_2$O$_4$.}
\label{fig:fig1}
\end{figure}

\begin{figure}[t]
\centering
\includegraphics[keepaspectratio,width=\linewidth]{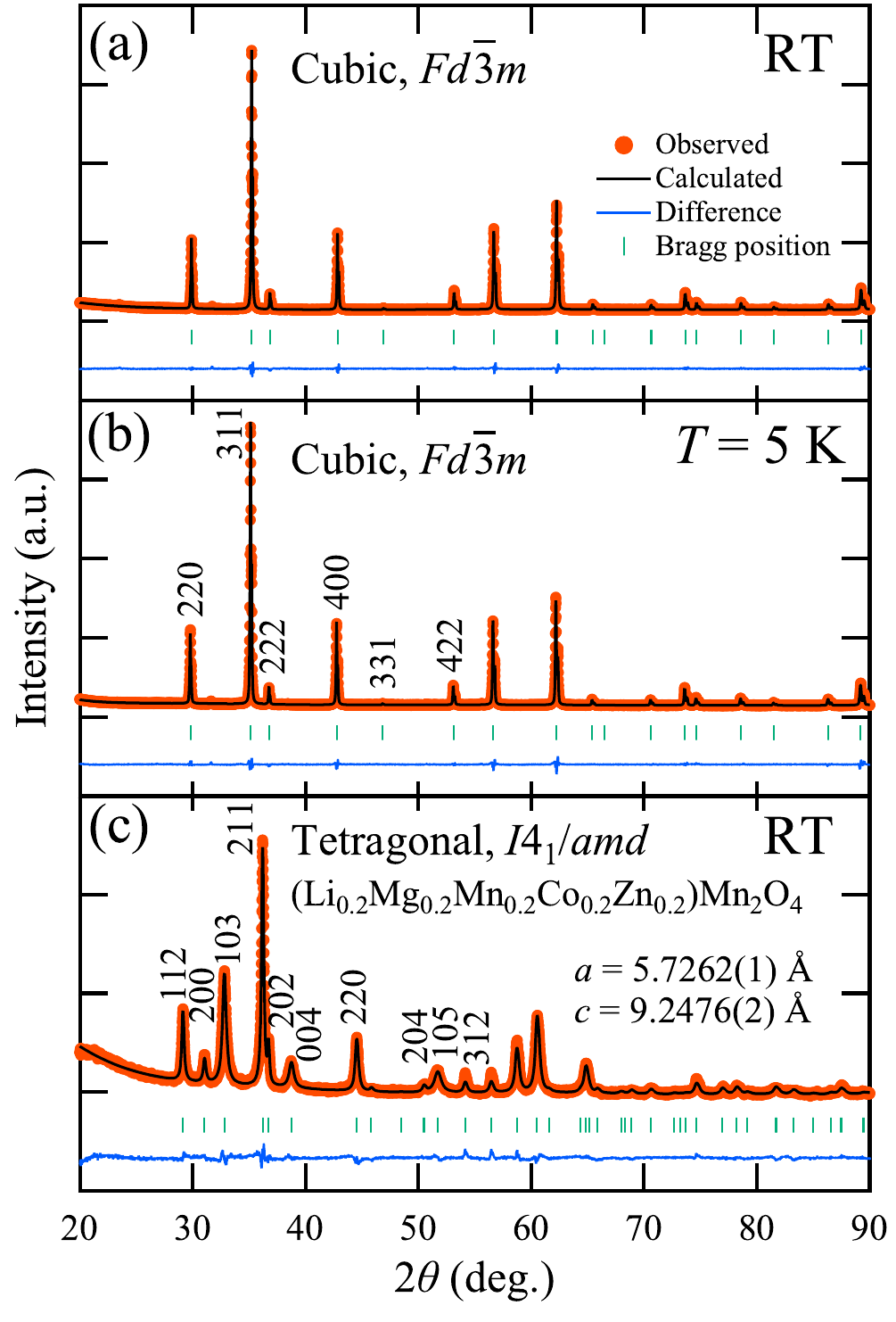}
\caption{Laboratory XRD patterns and Rietveld  refinements.  (a) Room-temperature XRD and (b) 5 K XRD patterns of (Li$_{0.2}$Mg$_{0.2}$Mn$_{0.2}$Co$_{0.2}$Zn$_{0.2}$)V$_2$O$_4$, refined assuming the cubic space group $Fd\bar{3}m$. No peak splitting indicative of symmetry lowering from the cubic spinel structure is observed down to 5 K. (c) Room-temperature XRD pattern of (Li$_{0.2}$Mg$_{0.2}$Mn$_{0.2}$Co$_{0.2}$Zn$_{0.2}$)Mn$_2$O$_4$, refined assuming the tetragonal space group $I4_1/amd$.}
\label{fig:fig2}
\end{figure}

\begin{figure}[t]
\centering
\includegraphics[keepaspectratio,width=\linewidth]{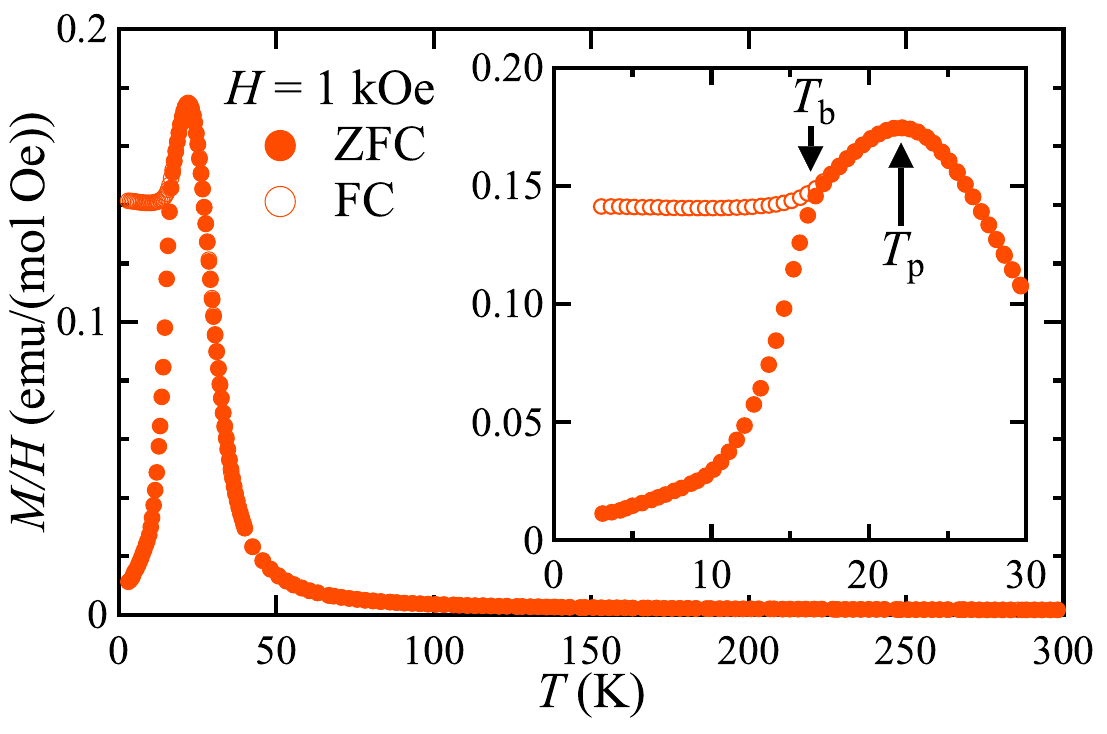}
\caption{Temperature dependence of the ZFC and FC magnetization $M(T)$ of (Li$_{0.2}$Mg$_{0.2}$Mn$_{0.2}$Co$_{0.2}$Zn$_{0.2}$)V$_2$O$_4$ measured under an applied field of 1 kOe. The inset shows an expanded view below 30 K. The values $T_{\rm p}$ and $T_{\rm b}$ represent the magnetization peak temperature and the ZFC-FC bifurcation temperature, respectively.}
\label{fig:fig3}
\end{figure}

\begin{figure}[t]
\centering
\includegraphics[keepaspectratio,width=\linewidth]{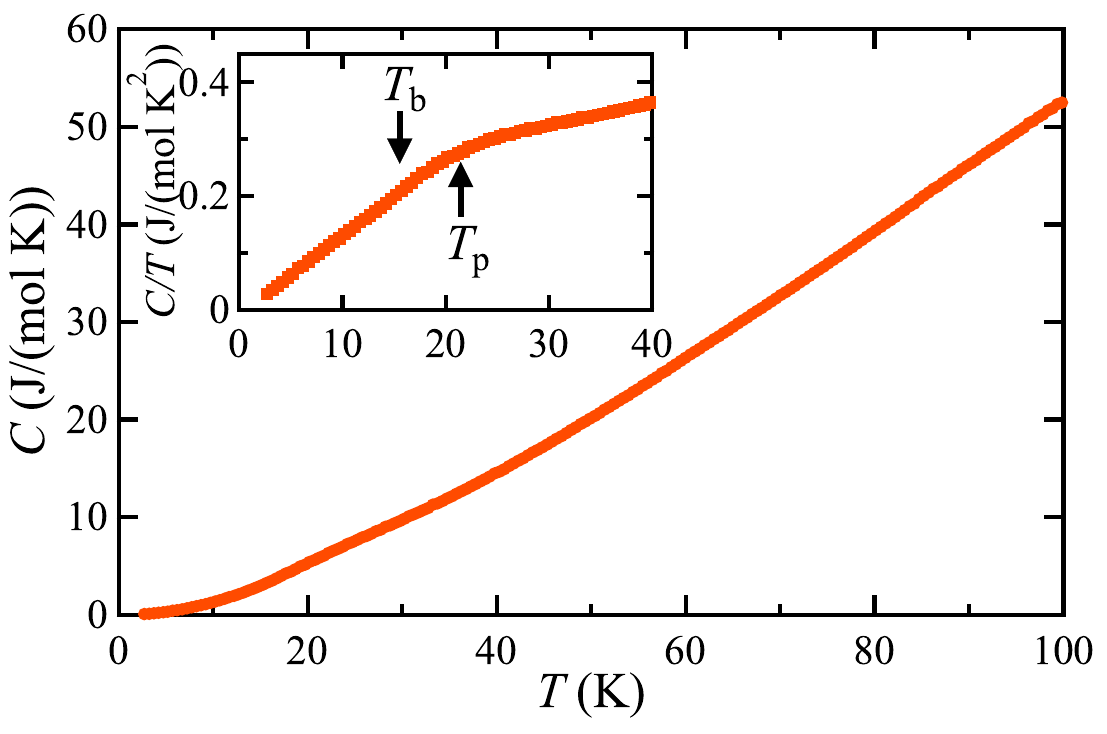}
\caption{Temperature dependence of the specific heat $C(T)$ of (Li$_{0.2}$Mg$_{0.2}$Mn$_{0.2}$Co$_{0.2}$Zn$_{0.2}$)V$_2$O$_4$ measured in zero magnetic field. The inset shows $C/T$ below 40 K. No anomaly associated with any phase transition is observed down to 3 K. The arrows indicate the magnetization peak temperature $T_{\rm p}$ and the ZFC-FC bifurcation temperature $T_{\rm b}$ shown in Fig. \ref{fig:fig3}.}
\label{fig:fig4}
\end{figure}

\begin{figure}[t]
\centering
\includegraphics[keepaspectratio,width=\linewidth]{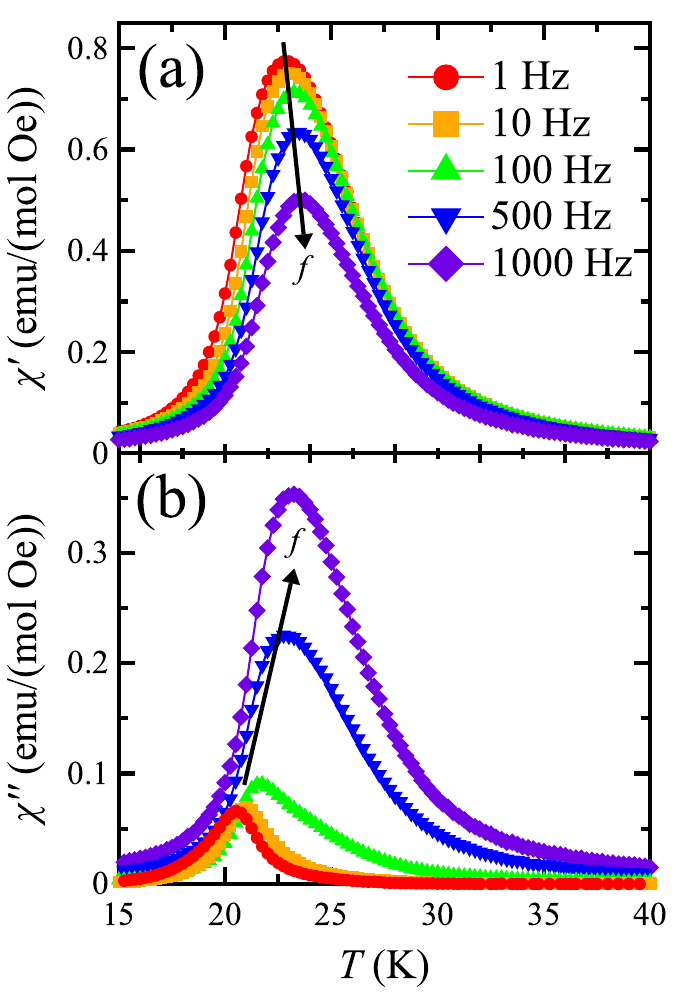}
\caption{Temperature dependence of the ac magnetic susceptibility of (Li$_{0.2}$Mg$_{0.2}$Mn$_{0.2}$Co$_{0.2}$Zn$_{0.2}$)V$_2$O$_4$ measured at several frequencies (1 Hz--1000 Hz) with ac excitation field of 2 Oe and zero dc magnetic field. (a) Real part $\chi'(T)$ and (b) imaginary part $\chi''(T)$. The arrows indicate the trend with increasing frequency, showing the frequency-dependent shift of the peak temperatures $T_f$ characteristic of glassy magnetic dynamics.}
\label{fig:fig5}
\end{figure}

\begin{figure*}[t]
\centering
\includegraphics[keepaspectratio,width=0.49\linewidth]{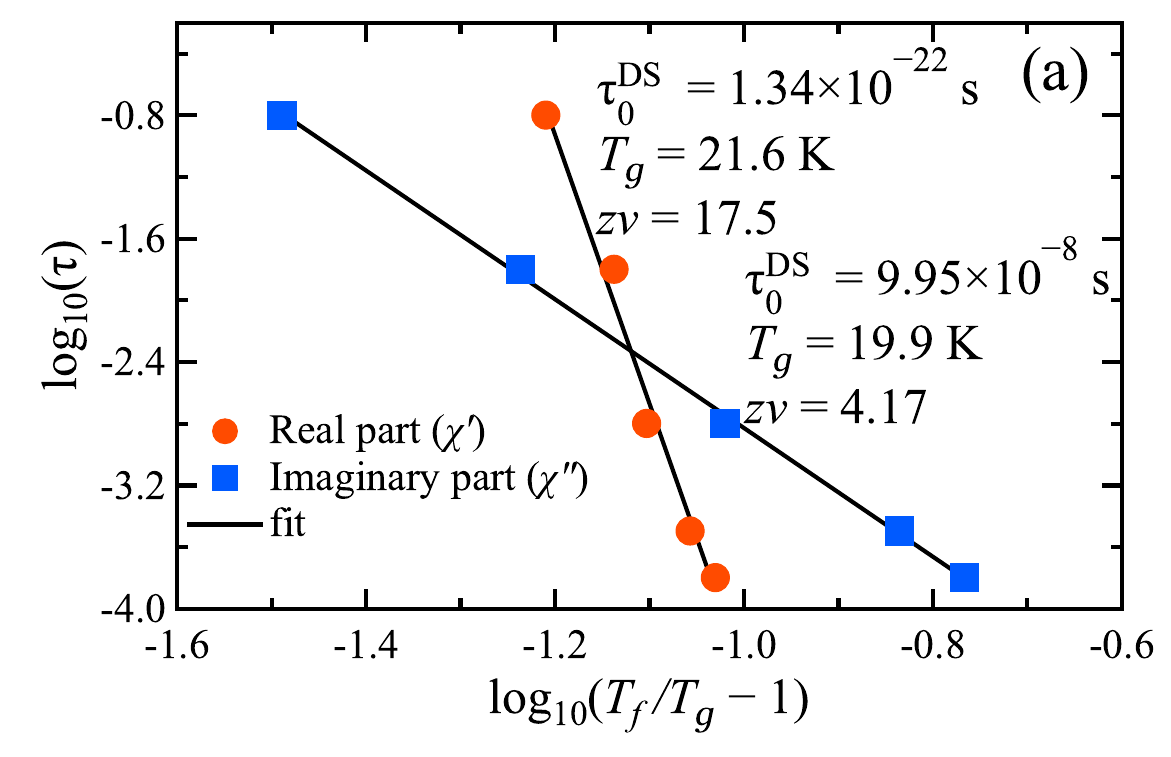}
\includegraphics[keepaspectratio,width=0.49\linewidth]{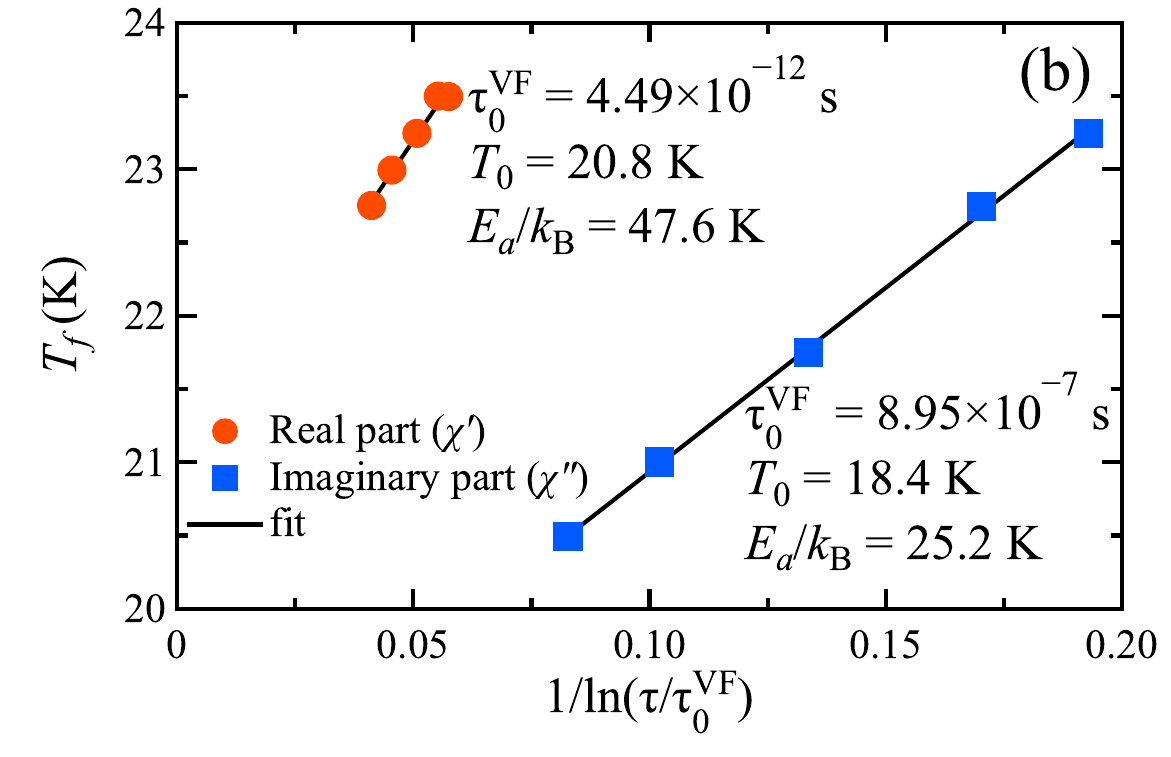}
\caption{Frequency dependence of the freezing temperature $T_f$ of (Li$_{0.2}$Mg$_{0.2}$Mn$_{0.2}$Co$_{0.2}$Zn$_{0.2}$)V$_2$O$_4$ analyzed using dynamic scaling and the Vogel--Fulcher model. (a) Dynamic scaling (power-law) analysis results obtained from a plot of ($\log_{10}(\tau)$) versus ($\log_{10}(T_f/T_g-1)$) [Eq. (3)]. (b) Vogel--Fulcher analysis results obtained from a plot of $T_f$ versus $1/\ln(\tau/\tau_0^{VF})$ [Eq. (5)]. Filled circles and squares represent the results obtained for the real ($\chi'$) and imaginary ($\chi''$) components of the ac susceptibility, respectively [Fig. \ref{fig:fig5}]. Solid lines are the corresponding least-squares fits.}
\label{fig:fig6}
\end{figure*}

\section{Results}
\subsection{Crystal structure}

Figures 1(a) and 1(b) present the synchrotron XRD patterns of (Li$_{0.2}$Mg$_{0.2}$Mn$_{0.2}$Co$_{0.2}$Zn$_{0.2}$)V$_2$O$_4$ measured at 300 K and 100 K, respectively, along with the Rietveld refinement results based on a cubic spinel structure with the space group $Fd\bar{3}m$. The observed patterns are well reproduced by the calculated profiles at both temperatures with no detectable impurity phases within the experimental resolution. Upon cooling to 100 K, the $a$-axis length contracts by approximately 0.095\% relative to that at 300 K. Furthermore, no peak splitting or additional reflections indicative of symmetry lowering are observed at temperatures down to 100 K. Figure 1(c) presents the SEM image together with EDS elemental mappings. All detectable constituent elements are homogeneously distributed over the sample, indicating a chemically uniform high-entropy phase without detectable elemental segregation. Note that Li is not included in the EDS mapping because of its low atomic number.

To further examine possible structural changes at lower temperatures, laboratory XRD measurements were performed at temperatures down to 5 K. As shown in Figs. 2(a) and 2(b), the diffraction patterns obtained at room temperature [Fig. 2(a)] and 5 K [Fig. 2(b)] are both successfully refined assuming a cubic spinel structure with space group $Fd\bar{3}m$, indicating the absence of any detectable symmetry lowering within the present experimental resolution. For comparison, Fig. 2(c) depicts the room-temperature laboratory XRD pattern of the high-entropy spinel manganate (Li$_{0.2}$Mg$_{0.2}$Mn$_{0.2}$Co$_{0.2}$Zn$_{0.2}$)Mn$_2$O$_4$. The diffraction pattern is well reproduced by the tetragonal space group $I4_1/amd$. This tetragonal structure is identical to that in the non-high-entropy spinel manganate ZnMn$_2$O$_4$, for which the cubic-to-tetragonal Jahn--Teller transition occurs at 1323 K \cite{Irani}.  The observations in Figs. 2(a)--2(c) strongly suggest that, for (Li$_{0.2}$Mg$_{0.2}$Mn$_{0.2}$Co$_{0.2}$Zn$_{0.2}$)V$_2$O$_4$, structural symmetry lowering associated with long-range orbital ordering is absent within the present experimental resolution.

\subsection{DC magnetization and specific heat}

Figure 3 depicts the temperature dependence of the dc magnetization $M(T)$ for (Li$_{0.2}$Mg$_{0.2}$Mn$_{0.2}$Co$_{0.2}$Zn$_{0.2}$)V$_2$O$_4$ measured under an applied magnetic field of 1 kOe. Upon cooling, the magnetization exhibits a ferrimagnetic-like increase below $\sim$50 K and a maximum at $T_{\rm p} \sim$ 22 K. Upon further cooling, the zero-field-cooled (ZFC) and field-cooled (FC) magnetization curves bifurcate below $T_{\rm b} \sim$ 17 K, as highlighted in the inset of Fig. 3. The ZFC magnetization also exhibits a slight kink at approximately $T_{\rm b}$. Figure 4 depicts the temperature dependence of the specific heat $C(T)$ for (Li$_{0.2}$Mg$_{0.2}$Mn$_{0.2}$Co$_{0.2}$Zn$_{0.2}$)V$_2$O$_4$. No sharp anomaly associated with a phase transition is observed at temperatures down to 3 K. As shown in the inset of Fig. 4, the $C/T$ curve exhibits only a small change in slope at approximately $T_{\rm p}$, while no discernible anomaly is observed at $T_{\rm b}$.

\subsection{AC susceptibility}

Figures 5(a) and 5(b) respectively present the temperature dependence of the real ($\chi'$) and imaginary ($\chi''$) parts of the ac susceptibility for (Li$_{0.2}$Mg$_{0.2}$Mn$_{0.2}$Co$_{0.2}$Zn$_{0.2}$)V$_2$O$_4$ measured at several frequencies in the range of 1 Hz--1000 Hz. As shown in Fig. 5(a), the peak temperature of the real component $\chi'$ gradually shifts higher with increasing frequency, and the peak height decreases gradually. In contrast, the imaginary component $\chi''$ shown in Fig. 5(b) exhibits a much stronger frequency dependence. The peak temperature shifts to higher temperatures with increasing frequency, accompanied by a pronounced increase in the peak intensity. The distinct frequency dependence observed in both $\chi'$ and $\chi''$ indicates a slow magnetic relaxation characteristic of glassy magnetic dynamics.

\begin{table*}[t]
\centering
\caption{\label{tab:table1} Dynamic parameters obtained from analyses of the frequency dependence of the ac magnetic susceptibility of (Li$_{0.2}$Mg$_{0.2}$Mn$_{0.2}$Co$_{0.2}$Zn$_{0.2}$)V$_2$O$_4$ [Fig. \ref{fig:fig5}]. The parameters derived from the real ($\chi'$) and imaginary ($\chi''$) components are listed separately. The Mydosh parameter $S$, dynamic scaling parameters ($\tau_0^{\rm DS}$, $T_g$, and $z\nu$) [Fig. 6(a)], and Vogel--Fulcher parameters ($\tau_0^{\rm VF}$, $T_0$, and $E_a/k_{\rm B}$) [Fig. 6(b)] are summarized.}
\begin{ruledtabular}
\begin{tabular}{cccccccc}
\textbf{Component}&Mydosh parameter $S$&$\tau_0^{\rm DS}$ (s)&$T_g$ (K)&$z\nu$&$\tau_0^{\rm VF}$ (s)&$T_0$ (K)&$E_a/k_{\rm B}$ (K)\\
\hline
\bm{$\chi'$}&0.011&1.34$\times$10$^{-22}$&21.6&17.5&4.49$\times$10$^{-12}$&20.8&47.6\\
\hline
\bm{$\chi''$}&0.045&9.95$\times$10$^{-8}$&19.9&4.17&8.95$\times$10$^{-7}$&18.4&25.2\\
\end{tabular}
\end{ruledtabular}
\end{table*}

\subsection{Dynamic magnetic relaxation}

To quantitatively characterize the glassy magnetic dynamics, we analyze the frequency $f$ dependence of the freezing temperature $T_f$ using the Mydosh parameter, dynamic scaling (power-law), and Vogel--Fulcher models \cite{Mydosh}. Here, the peak temperatures of $\chi'(T)$ [Fig. 5(a)] and $\chi''(T)$ [Fig. 5(b)] are defined as $T_f$.

The Mydosh parameter $S$ is the relative shift in freezing temperature $T_f$ per decade of frequency $f$, which is written as \cite{Mydosh},
\begin{equation}
S = \frac{\Delta T_f}{T_f\Delta\log_{10}f}.
\label{eq:Mydosh}
\end{equation}
Here $\Delta T_f$ is the change in $T_f$ for a  change of the logarithmic frequency ($\Delta\log_{10}f$), and $T_f$ is the freezing temperature at the lowest measured frequency of 1 Hz in the present study. From the experimental results shown in Fig. 5, the Mydosh parameter is determined to be $S=0.011$ for $\chi'$, and $S=0.045$ for $\chi''$.

In spin-glass systems, the frequency $f$ dependence of the freezing temperature $T_f$ is often analyzed using the critical slowing-down (power-law) relation derived from dynamic scaling theory \cite{Mydosh}:
\begin{equation}
\frac{\tau}{\tau_0^{\rm DS}} = \left(\frac{T_f}{T_g}-1\right)^{-z\nu},
\label{eq:Power}
\end{equation}
where $\tau$ is the relaxation time corresponding to the measured frequency ($\tau = 1/2\pi f$), $\tau_0^{\mathrm{DS}}$ is the characteristic relaxation time of a single spin flip of the fluctuating entities, $T_g$ is the spin freezing temperature in the limit as frequency approaches zero, and $z\nu$ is the dynamic exponent. To fit the experimental data [Fig. 5], Eq. (2) is rewritten as
\begin{equation}
\log_{10}(\tau) = \log_{10}(\tau_0^{\rm DS})-z\nu\log_{10}\left(\frac{T_f}{T_g}-1\right).
\label{eq:Power}
\end{equation}
 Figure 6(a) depicts the power-law fits based on Eq. (3). The fitting yields $\tau_0^{\rm DS} =1.34\times10^{-22}$ s, $T_g=21.6$ K, and $z\nu =17.5$ for $\chi'$, whereas $\tau_0^{\rm DS} =9.95\times10^{-8}$ s, $T_g=19.9$ K, and $z\nu =4.17$ for $\chi''$.

The phenomenological Vogel--Fulcher law is another model frequently used to describe slow magnetic dynamics in spin-glass systems, taking into account interactions among fluctuating entities \cite{Mydosh}:
\begin{equation}
\tau=\tau_0^{\rm VF}\exp\left[\frac{E_a}{k_B(T_f-T_0)}\right],
\label{eq:Power}
\end{equation}
where $\tau_0^{\rm VF}$ is the characteristic relaxation time in the Vogel--Fulcher model, $E_a$ is the activation energy, and $T_0$ is the empirical Vogel--Fulcher temperature, which is often interpreted as the interaction strength among the dynamic entities. To fit the experimental data [Fig. 5], we rewrite Eq. (4) as
\begin{equation}
T_f=T_0+\frac{E_a}{k_B}\frac{1}{\ln(\tau/\tau_0^{\rm VF})},
\label{eq:Power}
\end{equation}
Figure 6(b) presents the results of the Vogel--Fulcher analysis using Eq. (5). The evaluated values are $\tau_0^{\rm VF}=4.49\times10^{-12}$ s, $T_0=$20.8 K, and $E_a/k_B=$47.6 K for $\chi'$, and $\tau_0^{\rm VF}=8.95\times10^{-7}$ s, $T_0=$18.4 K, and $E_a/k_B=$25.2 K for $\chi''$. 

In Table I, the fitting parameter values obtained in this section for the real component $\chi'$ and imaginary component $\chi''$ are summarized. These analyses reveal markedly different dynamical parameters for the real and imaginary components, suggesting that the two components probe different aspects of magnetic relaxation.

\subsection{Cole--Cole analysis}

To further investigate the magnetic relaxation processes, Cole--Cole plots showing $\chi''$ versus $\chi'$ at different temperatures are constructed from the ac susceptibility measurements performed over an extended set of frequencies (1 Hz--1000 Hz), including those shown in Fig. 5 \cite{Topping}. Figure 7 compares the experimental Cole--Cole plots with the single Debye relaxation model at several temperatures (30 K--15 K). For an ideal single Debye relaxation process, the Cole--Cole relation is
\begin{equation}
\chi''(\chi')=\sqrt{(\chi'-\chi_S)(\chi_T-\chi')},
\label{eq:Cole}
\end{equation}
where $\chi_T$ and $\chi_S$ denote the isothermal and adiabatic susceptibilities in the limits of frequency $f\to$ 0 and $f\to\infty$, respectively. Eq. (6) predicts a semicircular Cole--Cole plot for a single relaxation process [the solid curves in Fig. 7]. As shown in Fig. 7, the experimental Cole--Cole plots exhibit temperature dependence as the temperature is lowered from 30 K to 15 K. At 30 K, the experimental data are well described by the single Debye model, indicating a single dominant relaxation process. Similarly good agreement is obtained at 23 K. Below 23 K, however, systematic deviations from the Debye model develop. The deviations become most pronounced at approximately 21 K and 19 K, where an additional relaxation component appears on the low-frequency side of the Cole--Cole plots. Upon further cooling to 17 K and 15 K, the deviation becomes smaller again while remaining discernible. These results indicate that the magnetic relaxation process evolves from one dominated by a single process at high temperatures to a more heterogeneous relaxation process in the vicinity of the magnetic freezing regime.

\begin{figure*}[t]
\centering
\includegraphics[keepaspectratio,width=\linewidth,trim= 0 120 0 120]{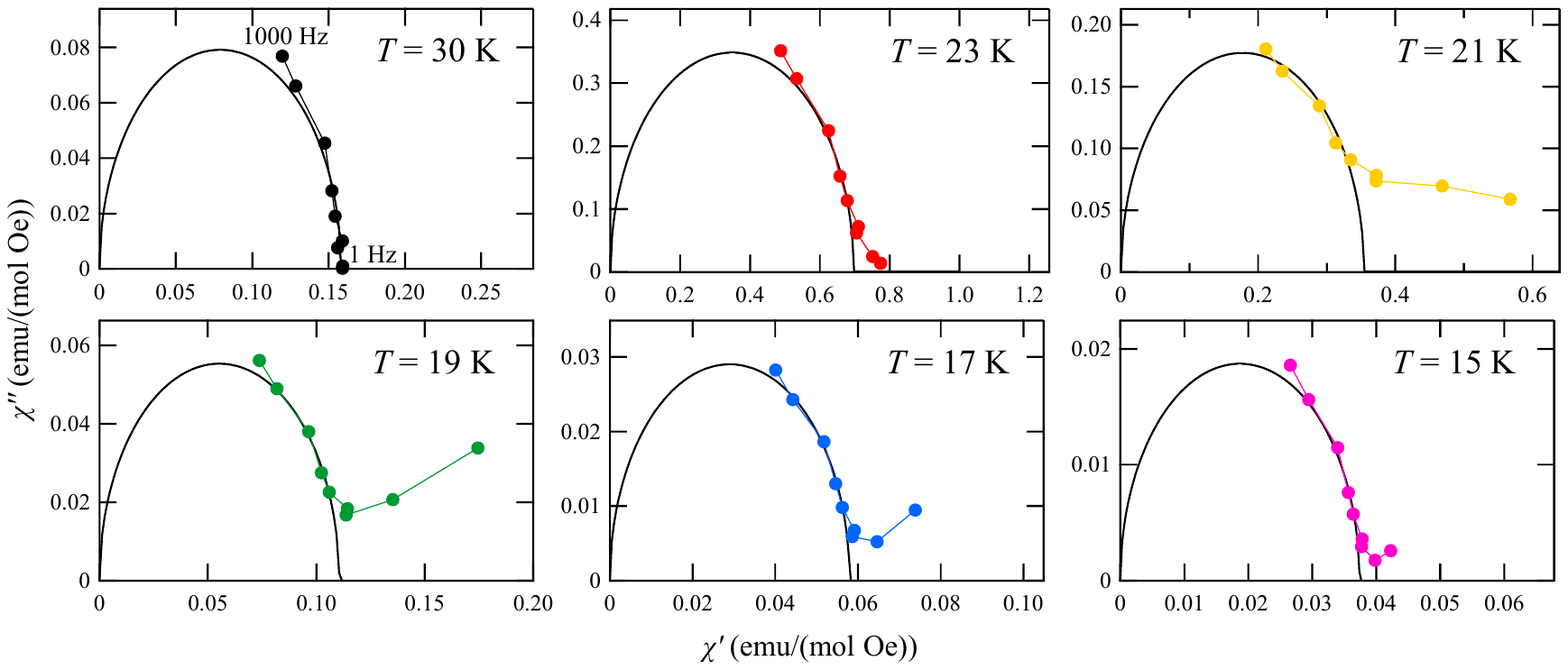}
\caption{Cole--Cole plots of the ac magnetic susceptibility measured at several temperatures (30 K--15 K) for (Li$_{0.2}$Mg$_{0.2}$Mn$_{0.2}$Co$_{0.2}$Zn$_{0.2}$)V$_2$O$_4$. Filled circles represent the experimental data constructed from the ac susceptibility measurements performed over an extended set of frequencies (1 Hz--1000 Hz), including those shown in Fig. \ref{fig:fig5}, while the solid curves are fits based on the single Debye relaxation model [Eq. (\ref{eq:Cole})]. Eq. (\ref{eq:Cole}) predicts a perfect semicircle for a single relaxation time. The experimental data at 30 K and 23 K are well described by a single relaxation process, whereas systematic deviations from the Debye model emerge below approximately 23 K and become most pronounced at approximately 21 K and 19 K, indicating the emergence of heterogeneous magnetic relaxation dynamics near the freezing regime.}
\label{fig:fig7}
\end{figure*}

\section{Discussion}
\subsection{Suppression of orbital ordering}

A central result of the present study is the absence of any detectable structural symmetry lowering in (Li$_{0.2}$Mg$_{0.2}$Mn$_{0.2}$Co$_{0.2}$Zn$_{0.2}$)V$_2$O$_4$ for temperatures down to 5 K.  This behavior is in marked contrast to that of conventional spinel vanadates $A$V$_2$O$_4$.  In $A$V$_2$O$_4$ with nonmagnetic $A$-site ions such as Zn, Mg, and Cd, a cubic-to-tetragonal or related symmetry-lowering structural transition is observed upon cooling, followed by magnetic ordering at lower temperatures \cite{Ueda,Reehuis,Lee2,Zhang,Mamiya,Wheeler,Nishiguchi,Onoda,Giovannetti,Watanabe}. This structural transition has generally been associated with long-range ordering of the V $t_{2g}$ orbitals, which lifts the orbital degeneracy and partially relieves the magnetic frustration in the pyrochlore sublattice. For magnetic $A$-site ions such as Mn and Fe, additional $A$--V exchange interactions further modify the spin-orbital-lattice coupled phase behavior, but structural transitions associated with orbital degrees of freedom remain a characteristic feature of these systems \cite{Adachi,Suzuki,Zhou,Chung,Garlea,Hardy,Nii1,Nii2,Gleason,Katsufuji,MacDougall,Zhang2,Kang,Kawaguchi}.  Therefore, the persistence of the cubic spinel structure at temperatures down to 5 K in the present high-entropy spinel vanadate indicates that long-range orbital ordering is strongly suppressed.

A comparison with the high-entropy spinel manganate (Li$_{0.2}$Mg$_{0.2}$Mn$_{0.2}$Co$_{0.2}$Zn$_{0.2}$)Mn$_2$O$_4$ is also instructive. As shown in Fig. 2(c), this compound exhibits a tetragonal structure at room temperature, consistent with the cooperative Jahn--Teller distortion commonly observed in spinel manganates $A$Mn$_2$O$_4$ with orbital-active Mn$^{3+}$ ions \cite{Irani}. Thus, configurational disorder on the $A$ site does not necessarily suppress cooperative orbital or Jahn--Teller ordering in spinel oxides.

A related phenomenon has been reported very recently in $A$-site-disordered high-entropy spinel chromites (h-$A$)Cr$_2$O$_4$ containing Ni and Cu \cite{Mandal}. In such systems, the intermediate tetragonal Jahn--Teller phase found in stoichiometric NiCr$_2$O$_4$ and CuCr$_2$O$_4$ is suppressed, whereas a cubic-to-orthorhombic structural transition and long-range antiferromagnetic ordering remain. This indicates that high configurational entropy does not simply eliminate cooperative symmetry-lowering transitions in spinel oxides, but can modify the balance between Jahn--Teller, magnetoelastic, and magnetic ordering tendencies. In contrast to the cases of (Li$_{0.2}$Mg$_{0.2}$Mn$_{0.2}$Co$_{0.2}$Zn$_{0.2}$)Mn$_2$O$_4$ and (h-$A$)Cr$_2$O$_4$, the absence of symmetry lowering in (Li$_{0.2}$Mg$_{0.2}$Mn$_{0.2}$Co$_{0.2}$Zn$_{0.2}$)V$_2$O$_4$ suggests that long-range orbital ordering on the V pyrochlore sublattice is destabilized by the combined effects of geometrical frustration and high-entropy-induced disorder.

It is useful to contrast the present results with those of recent studies on high-entropy perovskite vanadates (h-$R$)VO$_3$ \cite{Yan}. In such systems, long-range ordering of V $t_{2g}$ orbitals survives even in the presence of substantial rare-earth-site disorder, and the orbital state is governed mainly by the average ionic radius and the size variance of the rare-earth ions.  The present high-entropy spinel vanadate differs from (h-$R$)VO$_3$ in two important respects.  First, the V ions in the present compound form a geometrically frustrated pyrochlore lattice. Second, the high-entropy $A$ site in the present compound contains not only nonmagnetic ions but also magnetic Mn and Co ions.  Consequently, configurational disorder is expected to introduce not only local lattice randomness but also spatial randomness in the $A$--V magnetic exchange network.

This random exchange environment may play an important role in destabilizing the long-range orbital-ordered state.  In conventional spinel vanadates, orbital ordering and structural symmetry lowering are coupled to the formation of a more coherent spin-orbital-lattice state. In the present compound, however, each V tetrahedron experiences a different local environment through the random distribution of Li, Mg, Mn, Co, and Zn ions on the surrounding $A$ sites.  Such randomness can spatially modulate the local crystal field, the V--O bond geometry, and the magnetic exchange interactions involving the V spins. These effects are expected to compete with the cooperative interaction required for long-range orbital ordering. Possible charge disorder associated with charge compensation in the mixed-$A$-site composition, if present, may also contribute to the destabilization of a coherent orbital-ordered state.

We therefore suggest that the absence of a structural transition in (Li$_{0.2}$Mg$_{0.2}$Mn$_{0.2}$Co$_{0.2}$Zn$_{0.2}$)V$_2$O$_4$ does not simply reflect the removal of orbital degrees of freedom. Rather, the V $t_{2g}$ orbital degrees of freedom may remain active but fail to develop a long-range ordered pattern because of the spatially random spin-orbital-lattice environment created by the high-entropy $A$ site. In the present compound, the combination of geometrical frustration, configurational disorder, and random $A$--V exchange interactions is expected to destabilize a coherent spin-orbital-lattice state.  This picture suggests that the low-temperature magnetic dynamics discussed in the next subsection develop in a spatially nonuniform V-based electronic and magnetic environment.

\subsection{Glassy magnetic state and magnetic dynamics}

We next discuss the nature of the glassy magnetic state and its relaxation dynamics.  The dc magnetization exhibits a ferrimagnetic-like increase below $\sim$50 K and shows a maximum at $T_{\rm p} \sim$ 22 K, while the ZFC and FC curves bifurcate only below $T_{\rm b} \sim$ 17 K [Fig. 3].  In contrast, the specific heat shows no sharp anomaly associated with a thermodynamic phase transition for temperatures down to 3 K [Fig. 4].  These results indicate that the magnetic anomaly at approximately $T_{\rm p}$ is not a conventional long-range magnetic ordering transition. Rather, it is more naturally regarded as a crossover associated with the development and slowing down of V-sublattice magnetic correlations influenced by the random $A$--V exchange environment.

The separation between $T_{\rm p}$ and $T_{\rm b}$ is important for understanding the freezing process.  Although the magnetization starts to decrease below $T_{\rm p}$, the ZFC and FC magnetization curves remain nearly reversible between $T_{\rm p}$ and $T_{\rm b}$.  This indicates that the magnetic correlations developing at approximately $T_{\rm p}$ are not yet statically frozen on the time scale of the dc magnetization measurement.  Irreversible glassy freezing is established only at lower temperatures near $T_{\rm b}$, where the relaxation time becomes sufficiently long to break ergodicity on the dc time scale.  Thus, the present system appears to undergo a gradual freezing process rather than an abrupt magnetic phase transition.

The frequency dependence of the ac susceptibility supports this view [Fig. 5]. The Mydosh parameters are calculated to be $S = 0.011$ for $\chi'$ and $S = 0.045$ for $\chi''$ [Table I].  The former value is close to those typically observed in canonical spin glasses, whereas the latter is larger and closer to the range often discussed for cluster-glass-like systems \cite{Mydosh}. Therefore, the present compound cannot be classified simply as either a canonical spin glass or a conventional cluster glass.  It rather exhibits an intermediate dynamical character, suggesting that $\chi'$ and $\chi''$ probe different aspects of spin relaxation.

The dynamic scaling analysis also suggests that the freezing process is not described by a single homogeneous critical slowing-down process [Fig. 6(a)]. In particular, the fit to the peak temperatures of $\chi'$ yields an extremely short characteristic time, $\tau^{\rm DS}_0 = 1.34 \times 10^{-22}$ s, together with a large dynamic exponent, $z\nu = 17.5$ [Table I].  These values are difficult to interpret as physically meaningful microscopic spin-flip parameters \cite{Mydosh}. In contrast, the analysis of $\chi''$ gives $\tau^{\rm DS}_0 = 9.95 \times 10^{-8}$ s and $z\nu = 4.17$ [Table I], which are more consistent with the slow dynamics of correlated spin regions \cite{Mydosh}. Similarly, the Vogel--Fulcher analysis [Fig. 6(b)] yields markedly different parameters for $\chi'$ and $\chi''$ [Table I]. These differences also indicate that the real and imaginary components of the ac susceptibility probe different parts of a broad relaxation spectrum, rather than a single well-defined relaxation channel.

The Cole--Cole analysis provides more direct evidence for such a heterogeneous relaxation process.  As shown in Fig. 7, the Cole--Cole plots at 30 K and 23 K are approximately described by a single Debye relaxation process.  This suggests that, above the freezing regime, the magnetic response is dominated by a relatively narrow distribution of relaxation times.  Below approximately 23 K, however, systematic deviations from the single Debye behavior develop.  In particular, at approximately 21 K and 19 K, an additional component appears on the low-frequency side of the Cole--Cole plots.  This additional slow component indicates that the magnetic relaxation becomes heterogeneous near the freezing regime.

To visualize the evolution of this slow relaxation component more clearly, Fig. 8 shows the superposed Cole--Cole plots in the temperature range between 15 K and 23 K. The low-frequency branches become pronounced below approximately 23 K and form a broad common envelope in the freezing regime. This behavior indicates that the additional slow relaxation component develops systematically upon cooling below $\sim$23 K, rather than arising from an accidental deviation from the single Debye form. We therefore regard the low-frequency component as a signature of heterogeneous magnetic relaxation associated with the gradual freezing of spatially nonuniform V-based magnetic correlations.

\begin{figure}[b]
\centering
\includegraphics[keepaspectratio,width=\linewidth,trim= 0 0 0 0]{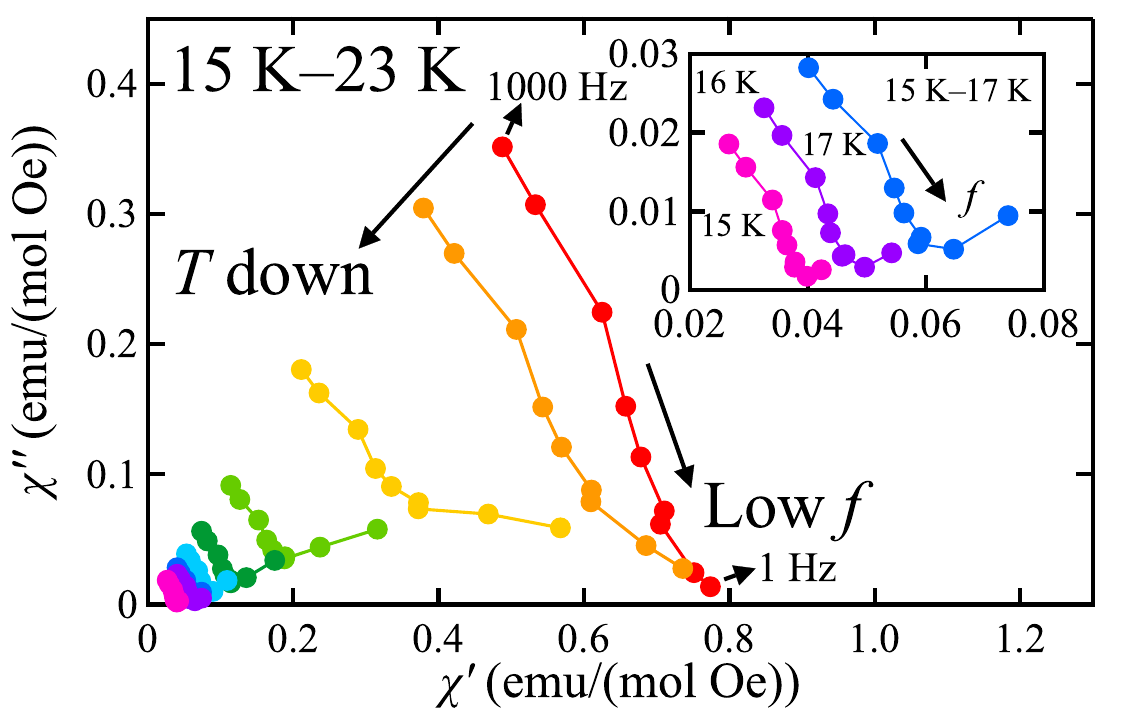}
\caption{Superposed Cole--Cole plots of the ac susceptibility for (Li$_{0.2}$Mg$_{0.2}$Mn$_{0.2}$Co$_{0.2}$Zn$_{0.2}$)V$_2$O$_4$ in the temperature range between 15 K and 23 K [Fig. 7]. The inset shows an expanded view of the plots at 15 K, 16 K, and 17 K. The superposition highlights the systematic evolution of an additional low-frequency relaxation component in the freezing regime, suggesting heterogeneous relaxation of V-based magnetic correlations.}
\label{fig:fig8}
\end{figure}

Here, the term ``heterogeneous’’ does not imply a macroscopic phase separation into distinct magnetic phases.  Instead, it indicates that the relaxation times and sizes of correlated spin regions are broadly distributed.  Such a distribution is naturally expected in the present high-entropy spinel vanadate, where the random occupation of the $A$ site by Li, Mg, Mn, Co, and Zn ions produces a spatially nonuniform exchange environment for the V pyrochlore sublattice.  In particular, magnetic Mn and Co ions on the $A$ site can locally introduce $A$--V exchange interactions, whereas nonmagnetic Li, Mg, and Zn ions dilute or modify these exchange pathways.  These local perturbations are not expected to remain confined to Mn- or Co-rich regions, because the V ions form a connected pyrochlore network with frustrated V--V interactions. Consequently, the random $A$-site environment can affect magnetic correlations over a broader region of the V-based magnetic network.

We therefore suggest that the glassy magnetic state in (Li$_{0.2}$Mg$_{0.2}$Mn$_{0.2}$Co$_{0.2}$Zn$_{0.2}$)V$_2$O$_4$ emerges through a gradual freezing of V-sublattice magnetic correlations in a spatially nonuniform exchange environment. The observed ferrimagnetic-like magnetization reflects the influence of local $A$--V exchange interactions involving Mn and Co ions. These random $A$--V interactions, together with frustrated V--V interactions, are expected to shape the glassy dynamics as a bulk response of the connected V pyrochlore network. The magnetic dynamics exhibit both canonical-spin-glass-like and cluster-glass-like features, reflecting a broad distribution of relaxation times. This heterogeneous glassy behavior is consistent with the high-entropy-induced disorder discussed above and provides an important link between the suppression of long-range orbital ordering and the emergence of a glassy magnetic state.

\subsection{Spin-orbital state}

The results discussed above suggest that the low-temperature state of (Li$_{0.2}$Mg$_{0.2}$Mn$_{0.2}$Co$_{0.2}$Zn$_{0.2}$)V$_2$O$_4$ should be viewed as a spin-orbital state characteristic of a high-entropy frustrated spinel vanadate.  In conventional spinel vanadates $A$V$_2$O$_4$, the orbital degrees of freedom of V$^{3+}$ ions are coupled to structural symmetry lowering and magnetic ordering \cite{Ueda,Reehuis,Lee2,Zhang,Mamiya,Wheeler,Nishiguchi,Onoda,Giovannetti,Watanabe,Adachi,Suzuki,Zhou,Chung,Garlea,Hardy,Nii1,Nii2,Gleason,Katsufuji,MacDougall,Zhang2,Kang,Kawaguchi}. The resulting orbital-ordered state provides a coherent spin-orbital-lattice background, which partially relieves magnetic frustration in the V pyrochlore sublattice.  In the present high-entropy spinel vanadate, however, no detectable structural symmetry lowering is observed at temperatures down to 5 K, and no thermodynamic anomaly associated with a phase transition is observed at temperatures down to 3 K.  Thus, the system does not develop the conventional coherent spin-orbital-lattice ordered state found in ordinary $A$V$_2$O$_4$ compounds.

Instead, the present results suggest the formation of a spatially nonuniform V-based spin-orbital-lattice environment.  The random occupation of the $A$ site by Li, Mg, Mn, Co, and Zn ions can modulate the local crystal field, V--O bond geometry, charge balance, and $A$--V exchange interactions.  Such local perturbations are expected to affect both the orbital and spin degrees of freedom in the connected V pyrochlore network.  In this situation, the V $t_{2g}$ orbital degrees of freedom may remain relevant at low temperatures, but they do not develop a long-range ordered pattern.  The absence of long-range orbital ordering therefore does not necessarily imply a simple orbital-inactive state; rather, it may indicate that the orbital sector remains spatially disordered or fluctuating in the high-entropy environment.

This picture also provides a natural framework for understanding the glassy magnetic dynamics. For (Li$_{0.2}$Mg$_{0.2}$Mn$_{0.2}$Co$_{0.2}$Zn$_{0.2}$)V$_2$O$_4$, the ferrimagnetic-like magnetization reflects the influence of local Mn--V and Co--V exchange interactions. The frequency-dependent ac susceptibility and Cole--Cole analyses indicate that these random $A$--V exchange interactions, together with frustrated V--V interactions, lead to heterogeneous relaxation of V-based magnetic correlations. These magnetic dynamics are therefore more naturally regarded as the freezing of V-based magnetic correlations in a spatially nonuniform spin-orbital-lattice background, rather than as a simple spin glass on a rigid orbital-inactive lattice. We thus suggest that high configurational entropy destabilizes long-range orbital ordering and stabilizes a glassy magnetic state with heterogeneous relaxation dynamics in the frustrated V pyrochlore network. The present compound may therefore represent a high-entropy-induced spatially nonuniform spin-orbital state, in which long-range orbital order is suppressed while V-based magnetic correlations freeze heterogeneously.

\section{Summary}

In summary, we investigated the structural, thermodynamic, and magnetic properties of the high-entropy spinel vanadate (Li$_{0.2}$Mg$_{0.2}$Mn$_{0.2}$Co$_{0.2}$Zn$_{0.2}$)V$_2$O$_4$, in which the orbital-active V ions form a geometrically frustrated pyrochlore sublattice.

X-ray diffraction measurements revealed that the cubic spinel structure is preserved at temperatures down to 5 K without any detectable symmetry lowering.  This behavior is in marked contrast to conventional spinel vanadates $A$V$_2$O$_4$, where long-range ordering of V $t_{2g}$ orbitals is generally accompanied by a symmetry-lowering structural transition. Comparison with the high-entropy spinel manganate (Li$_{0.2}$Mg$_{0.2}$Mn$_{0.2}$Co$_{0.2}$Zn$_{0.2}$)Mn$_2$O$_4$, which exhibits a tetragonal structure at room temperature, further indicates that configurational disorder on the $A$ site does not necessarily suppress cooperative orbital or Jahn--Teller ordering in spinel oxides.  These results suggest that long-range orbital ordering on the V pyrochlore sublattice is strongly suppressed in the present high-entropy spinel vanadate.

Magnetic and thermodynamic measurements revealed a glassy magnetic state without conventional long-range magnetic ordering.  The dc magnetization exhibits a ferrimagnetic-like increase below $\sim$50 K, a maximum at $T_{\rm p} \sim$ 22 K, and a ZFC--FC bifurcation below $T_{\rm b} \sim$ 17 K, whereas the specific heat shows no sharp anomaly associated with a thermodynamic phase transition for temperatures down to 3 K.  The frequency dependence of the ac susceptibility demonstrates trends consistent with the slow magnetic relaxation characteristic of glassy dynamics.  Analyses based on the Mydosh parameter, dynamic scaling law, and Vogel--Fulcher law indicate that the magnetic dynamics cannot be described by a single homogeneous freezing process, but instead exhibit an intermediate character between canonical-spin-glass-like and cluster-glass-like behavior.

Cole--Cole analyses further revealed that the magnetic relaxation evolves from being dominated by a single Debye-like process at higher temperatures to a heterogeneous process near the freezing regime.  In particular, an additional low-frequency relaxation component develops below approximately 23 K and becomes pronounced around 21 K and 19 K.  This behavior suggests that the glassy magnetic freezing involves a broad distribution of relaxation times associated with spatially nonuniform V-based magnetic correlations.

We therefore suggest that (Li$_{0.2}$Mg$_{0.2}$Mn$_{0.2}$Co$_{0.2}$Zn$_{0.2}$)V$_2$O$_4$ realizes a high-entropy-induced spatially nonuniform spin-orbital state. In this state, long-range orbital ordering is suppressed, while the V $t_{2g}$ orbital degrees of freedom may remain locally relevant in a spatially disordered or fluctuating form.  The random occupation of the $A$ site introduces local lattice randomness and random $A$--V exchange interactions, which affect the frustrated V pyrochlore network and give rise to heterogeneous glassy magnetic dynamics. Our results suggest that high configurational entropy provides an effective route to destabilize long-range orbital ordering and to stabilize spatially heterogeneous glassy spin-orbital dynamics in frustrated spinel vanadates.

\begin{acknowledgments}
This work was partly supported by JSPS KAKENHI Grant No. JP21K03476. The synchrotron XRD experiments were performed at SPring-8 with the approval of the Japan Synchrotron Radiation Research Institute (JASRI) (Proposal No. 2025B1563).
\end{acknowledgments}

\end{document}